\documentclass[conference]{IEEEtran}
\usepackage{cite}
\usepackage{amsmath,amssymb,amsfonts}
\usepackage{amsthm}
\usepackage{algorithmic}
\usepackage{graphicx}
\usepackage{textcomp}
\usepackage{xcolor}
\usepackage{orcidlink}
\usepackage{hyperref}
\usepackage{cleveref}
\usepackage{comment}

\usepackage{multirow}
\usepackage{tikz}
\usetikzlibrary{automata,positioning,fit,shapes.geometric}
\usepackage{subcaption}
\usepackage{algorithm2e}
\usepackage{mathtools}

\usepackage{siunitx}
\usepackage{adjustbox}

\usepackage{booktabs}

\usepackage{mesmacros}
\usepackage{pgfplots}
\usepgfplotslibrary{units}
\pgfplotsset{compat=1.18}

\usepackage{enumitem}

\theoremstyle{definition}
\newtheorem{definition}{Definition}

\newtheorem*{remark*}{Remark}
\newtheoremstyle{noparentheses}
  {3pt}
  {3pt}
  {\normalfont}
  {}
  {\bfseries}
  {.}
  {.5em}
  {\thmname{#1}\thmnumber{ #2}\thmnote{ #3}} 

\theoremstyle{noparentheses}
\newtheorem{AEBSproperty}{}[section] 

\title{Refining Timing Uncertainty from Logical Time Specification to Operation\thanks{This work has been supported by the European Chips Joint Undertaking under Framework Partnership Agreement No 101139789 (HAL4SDV).}}

\author{
  \IEEEauthorblockN{
    Pavlo Tokariev\textsuperscript{1} \orcidlink{0000-0002-4603-4770}, 
    and Julien Deantoni\textsuperscript{1,2} \orcidlink{0000-0001-6962-7846}
  }
  \IEEEauthorblockA{
    \textsuperscript{1}\textit{Inria, Kairos Team}, Sophia-Antipolis, France \\
    \textsuperscript{2}\textit{Université Côte d'Azur, I3S, CNRS}, Nice, France \\
    Email: pavlo.tokariev@inria.fr, julien.deantoni@univ-cotedazur.fr
  }
}

\IEEEoverridecommandlockouts

\begin{document}
\maketitle

\begin{abstract}

Real-time and cyber-physical systems are developed through successive refinements from abstract requirements to platform deployments. While timing knowledge evolves throughout this process, existing stochastic real-time formalisms typically require uncertainty to be embedded from the outset or necessitate model reconstruction when new timing information becomes available, hindering iterative timing engineering.

This paper presents a refinement-oriented timing specification framework built upon the Clock Constraint Specification Language (\ccsl). It supports the progressive introduction of timing knowledge across three levels: logical timing relations, quantitative real-time constraints, and stochastic timing models. Instead of altering behavioural semantics, stochastic information is attached directly to timing quantities as a refinement. This enables implementation measurements and operational observations to be incorporated incrementally within a unified declarative framework.

We implement our approach in an OCaml-based simulation tool and evaluate it on a simplified but representative Software-Defined Vehicle use case.
\end{abstract}


\section{Introduction}
Real-time and cyber-physical systems are rarely designed in a single step. Instead, they are progressively refined from abstract requirements towards platform-specific implementations. This refinement process is particularly important for timing requirements, which initially express logical ordering relations between events and are later enriched with quantitative information regarding execution times, communication delays, scheduling effects, and platform characteristics. Logical Time approaches and synchronous design methodologies have long advocated the separation of functional and temporal concerns in support of such incremental development processes~\cite{berry1983esterel,caspi.etal_1987,mallet_2008oct}.

Recent trends such as Software Defined Vehicles (SDV) further reinforce the need for incremental timing engineering. In these systems, functionality is increasingly determined by software configurations rather than hardware architectures and may evolve throughout the system lifecycle with software updates. As implementations mature, platforms evolve, and measurements become available during validation and operation, knowledge about timing behaviour becomes progressively more precise. Timing specifications must therefore accommodate evolving information while preserving continuity across development stages.

Several stochastic real-time formalisms have been proposed to model timing uncertainty, including stochastic timed automata and associated analysis frameworks~\cite{bohnenkamp.etal_2006oct,bouyer.etal_2016,bulychev.etal_2012jul}. These approaches provide powerful modelling and verification capabilities but generally introduce probability directly into the behavioural models. As a consequence, uncertainty assumptions often need to be established early in the modelling process, and incorporating newly available timing information may require substantial modifications to existing models.

Similar observations can be made for probabilistic timing specification approaches, including probabilistic extensions of the Clock Constraint Specification Language (\ccsl)~\cite{du.etal_2018nov,huang.etal_2019nov}. In these approaches, probabilities are typically associated with events or constraint satisfaction. While this provides useful mechanisms for analysing uncertain behaviour, it changes the interpretation of timing constraints from strict behavioural restrictions to probabilistic properties. As a result, the progressive introduction of timing uncertainty does not naturally align with the classical refinement intuition, where additional knowledge restricts the set of admissible behaviours.

As the contribution of this paper, we investigate how timing uncertainty can itself be treated as a refinement artifact. Our objective is not merely to introduce stochastic constructs into \ccsl~\cite{mallet_2008oct,deantoni.etal_2014nov}, but to provide a continuous timing-specification process through which timing knowledge can evolve together with the system. We build upon \ccsl and first refine logical timing specifications into quantitative real-time specifications by relating logical instants to physical time. We then introduce a stochastic refinement layer that allows timing quantities to be constrained by probability distributions.

The key idea of our approach is to associate uncertainty with timing quantities rather than with behavioural semantics or constraint satisfaction. This enables timing specifications to evolve through three successive refinement levels: logical timing specifications, quantitative real-time specifications, and stochastic timing specifications. Each level enriches the previous one with additional timing knowledge while preserving a common declarative foundation. In this sense, refinement follows the classical intuition of trace refinement: each step introduces additional timing information and further constrains the admissible behaviours represented by the specification.
Consequently, stochastic timing models may be derived not only from the initial design assumptions but also from measurements later collected during implementation, validation, or operation, without requiring a change of the modelling paradigm.
To evaluate the approach, we implement the framework as a Monte Carlo simulation environment and apply it to a SDV case study involving an Advanced Driver-Assistance System function. Starting from architectural timing specifications and functional chains, we progressively refine timing information and obtain stochastic timing models that support the analysis of end-to-end reaction times.



The paper is structured as follows: \Cref{sec:running} introduces the running example, and \Cref{sec:preliminaries} reviews \ccsl semantics. Our core contribution--a stochastic extension to a revisited \rtccsl language--is defined in \Cref{sec:contribution} and evaluated on a reaction time analysis in \Cref{sec:evaluation}. Finally, \Cref{sec:discussion} discusses the refinement process, and \Cref{sec:conclusion} concludes with future work.

\section{Running example}
\label{sec:running}

SDV architectures are a representative application domain for timing refinement. By introducing a Hardware Abstraction Layer (HAL) and standardized interfaces such as the Vehicle Signal Specification, software functions become largely decoupled from the underlying hardware platform. While this improves portability and enables software evolution through over-the-air updates, it also introduces additional uncertainty regarding execution, communication, and access latencies. Consequently, timing properties can no longer be entirely characterized from the initial architecture and must be progressively refined as implementation and operational information becomes available.

To illustrate the proposed refinement process, we use a simplified Autonomous Emergency Braking System (AEBS). AEBS is a safety-critical vehicular function that monitors the environment and triggers braking actions when a collision risk is detected. The considered architecture 
comprises three components: a radar sensor (\textbf{s}), a controller (\textbf{c}), and a braking actuator (\textbf{a}) interacting through the HAL.


AEBS, whose correctness depends not only on functional behaviour but also on reaction time, is used throughout the paper to illustrate the refinement process from logical requirements to operational timing models.

\section{Preliminaries}
\label{sec:preliminaries}


The Clock Constraint Specification Language~\cite{mallet_2008oct} is a language of constraints as statements over logical clocks, which are totally ordered sets of ticks.
Each constraint expresses a restriction on the relative order of ticks of its argument clocks. Such constraints allow expressing every safety property that can be expressed by the Property Specification Language (IEEE standard~\cite{PSL})~\cite{gascon:hal-00597086}.
Modelling-wise, a clock represents an identifiable family of events in the modelled system and constructs a multiform time behaviour of the said system.
In the following, we provide the definition of the denotational semantics of the language~\cite{deantoni.etal_2014nov}.

\begin{definition}[Time structure]
  \label{def:time-structure}
  A time structure is a tuple $\langle I , \equiv_I , \prec_I \rangle$, where \( I \) is a set of ticks, equivalence relation \( \equiv_I \) representing tick coincidence, i.e. assignment of ticks to the same logical steps, and strict partial order relation \( \prec_I \) that denotes precedence (``happens before'' relation~\cite{lamport_1978jul}).
  The derived preorder $a \preccurlyeq_I b \defeq a \prec_I b \lor a \equiv_I b$ represents causality relation.
\end{definition}

\begin{definition}[Logical clock]
  \label{def:denot-clock}
  A logical clock $\lc{c}$ is an index function $[0,i^{\max}) \to I$, mapping indexes to ticks, with indices always preceding the associated maximum index \( i^{\max} \in \N \cup\{+\infty\} \).
  The function is strictly monotonic \( \forall i \in [1,i^{\max}): c(i-1) \prec_I c(i) \), to enforce the total order in a logical clock.
\end{definition}

We denote the set of all clocks as \( C \).
We define testing that a tick with index $i$ exists in a clock $c \in C$ as $i \in \dom{c} \defeq (0 \leq i < i_c^{\max})$, where $i_c^{\max}$ is the $ i^{\max}$ of the clock $c$.
The indexing expression of a clock is its application $c[i] \defeq c(i)$, where $c \in C$ and $0 \leq i < i_c^{\max} \in \N$ is assumed to hold.

We define the set of all possible constraints over any clocks as \(\constr\), and the set of all specifications as the power set of the constraints \(\spec = \pow(\constr)\).
A \ccsl constraint semantics is then defined as a predicate \(P(\langle I , \equiv_I , \prec_I \rangle)\) in first-order logic over some time structure \(\langle I , \equiv_I , \prec_I \rangle\).

The relevant list of the constraint definitions is given in \Cref{tab:index_semantics}. Other constraint definitions can be found in~\cite{deantoni.etal_2014nov}.
In order to shorten the definitions of the \ccsl constraints, it is assumed that when expression involving $c[i]$ is written, like $a[i] \Join b[j]$, it is interpreted as $(i \in \dom{a} \land j \in \dom{b}) \implies \left(a[i] \Join b[j]\right)$, where \( \Join \in \{\prec_I, \preccurlyeq_I\} \).
Effectively meaning that when a tick does not exist, the relation with the expression is ``skipped'' in the predicate, allowing the definitions to use indexes over whole \(\N\), even in the case of finite clocks.
While in most of the constraints it is the only expected behaviour, in some, like precedence, additional conditions on tick existence are added explicitly.
In the case of usage of the coincidence relation \( a[i] \equiv_I b[j] \), existence of one tick implies existence of the other, i.e. this relation rewrites into $(i\in \dom{a} \iff j \in \dom{b}) \implies a[i] \equiv_I b[j]$.

Naturally, the semantics of a specification is defined as a conjunction of its constraint predicates on the same time structure.
Thus, solving a specification is equivalent to checking existence of a time structure that satisfies all predicates and is a valid time structure according to \Cref{def:time-structure}, i.e.:
\begin{equation}
    \exists \langle I , \equiv_I , \prec_I \rangle: \bigwedge^n_i P_i(\langle I , \equiv_I , \prec_I \rangle)
\end{equation}
A more concrete solution to a specification is a schedule (trace), defined as a function of steps to sets of ticked clocks $\sigma: \N \to \pow(C)$, and which provides a total order to the ticks of a time structure.
The empty schedule $\sigma_\emptyset : i \mapsto \sigma_\emptyset(i) = \emptyset$ satisfies any specification.
When only the empty schedule satisfies a specification, it is immediately deadlocked and so considered invalid.

\begin{table*}[!htb]
  \centering
  \caption{Denotational definitions of \ccsl constraints, $a,b,c,r \in C,\;d \in \N,p \in \N_{>0}$}
  \label{tab:index_semantics}
  \begin{adjustbox}{max width=\linewidth}
    \begin{tabular}{|l|l|l|}
      \hline
      Constraint    & Notation                           & Definition, $\forall i \in \N$                                                                                        \\
      \hline
      \multirow{6}{*}
      
      Causality     & $a \preccurlyeq b$                 & $a[i] \preccurlyeq_I b[i] \land i \in \dom{b} \implies i \in \dom{a}$                                                 \\
      Coincidence   & $a = b$                            & $a[i] \equiv_I b[i]$                                                                                                  \\
      Strict Alternation   & $a \strictalternates b$                  & $a[i] \prec_I b[i] \prec_I a[i+1]$                                                                                    \\
       Alternation   & $a \alternates b$                  & $a[i] \prec_I b[i] \preccurlyeq_I a[i+1]$                                                                                    \\
      \multirow{8}{*}
    
      Delay         & $b = a \;\$\; d$                   & $b[i] \equiv_I a[i + d]$ \\
      Sampling      & $c = \sampledon{a}{b}$             & $(\exists j \in \N: c[j] \equiv_I b[i]) \iff ((\exists k \in \N: b[i-1] \prec_I a[k] \preccurlyeq_I b[i]) \land i \in \dom{b}) $               \\
      \hline
    \end{tabular}
  \end{adjustbox}
\end{table*}

\section{Contribution}
\label{sec:contribution}

\ccsl has been used in various domains to make timing requirements explicit~\cite{goknil:hal-00850673,millo:hal-01644294,deantoni:hal-03374955,holtmann:hal-03375049}. Here we propose to use it to specify the safety properties to be preserved along the refinement.
The proposed refinements progressively constrain its interpretation in physical time: \rtccsl introduces real-time constraints reflecting some implementation details together with real-time mappings and timing bounds, while the stochastic extension further characterises these mappings through probability distributions. 
Unlike stochastic real-time approaches where uncertainty is embedded directly in behavioural or scheduling models, our uncertainty is attached to timing quantities that already appear in the specification. As a result, operational knowledge can be refined independently of the logical and real-time structure of the model. To support such refinements, the initial \ccsl specification should capture the logical timing intent while preserving sufficient flexibility for subsequent real-time and stochastic characterisation, a point we further discuss in \Cref{sec:discussion}.

To formalize this progressive design flow, we adopt the classic definition of refinement formalized by Hoare and He~\cite{Hoare2005}, where refinement is defined as the inclusion of all traces of a refined process in those of the process that it refines. However, as a specification is refined from abstract requirements to physical implementation, new internal mechanisms (such as intermediate communication events or middleware triggers) may introduce a set of internal clocks unique to that layer. 

To handle this, we define a classical projection operator $\upharpoonright_{C}$, which restricts a trace to a set of clocks $C$ by filtering out any ticks belonging to clocks outside $C$. Let $\mathcal{T}(S)$ denote the set of valid traces satisfying a specification $S \in \spec$, and let $C_{\text{abs}}$ be the set of clocks defined in the abstract specification \(S_{\text{abs}}\). A concrete specification $S_{\text{conc}}$ (defined over a larger clock set $C_{\text{conc}}$ where $C_{\text{abs}} \subseteq C_{\text{conc}}$) validly refines $S_{\text{abs}}$ if and only if its traces form a subset of the abstract traces, when projected onto the abstract clock set:
$ \{ \sigma \upharpoonright_{C_{\text{abs}}} \mid \sigma \in \mathcal{T}(S_{\text{conc}}) \} \subseteq \mathcal{T}(S_{\text{abs}}) $.

This definition allows the implementation to introduce necessary technical complexity while guaranteeing that the observable behaviour strictly preserves all original safety properties. We establish this projected trace inclusion across three execution layers:

\begin{description}[leftmargin=0.2cm, topsep=0pt, itemsep=0.5ex]    
    \item[Logical Traces ($\mathcal{T}_L$)]: At the most abstract level, a logical trace $\sigma \in \mathcal{T}_L$ is a total order of steps over the initial set of clocks $C_L$ that satisfies the underlying time structure and complies with the \ccsl constraints.
    \item[Real-Time Logical Traces ($\mathcal{T}_{RT}$)]: Moving to \rtccsl may introduce new clocks $C_{RT}$ (where $C_L \subseteq C_{RT}$) to model platform-specific artifacts like network latencies or hardware timers. A real-time logical trace complements the steps with physical timestamps
    . The refinement requires that the real-time traces, stripped of these new internal clocks, are valid logical traces: $\mathcal{T}_{RT} \upharpoonright_{C_L} \subseteq \mathcal{T}_L$. Since \rtccsl introduces only additional constraints over \ccsl traces, refinement follows directly from the monotonicity of constraint satisfaction: adding constraints can only reduce the set of admissible traces. The projection is required solely to account for the introduction of auxiliary clocks at lower refinement levels.
    
    \item[Stochastic Traces ($\mathcal{T}_{ST}$)]: The stochastic extension represents the final layer, but it fundamentally differs in nature from the previous step. Rather than further pruning the set of permissible execution paths, the stochastic layer operates on the exact same structural boundaries as the real-time layer. Instead of reducing the behavioural space, it overlays a probability measure $\mu$ across $\mathcal{T}_{RT}$: \(\mathcal{T}_{ST} = (\mathcal{T}_{RT}, \mu)\), by dictating how timing values are sampled from the continuous real-time intervals during execution, shifting the focus from possibilistic correctness to probabilistic evaluation. 
\end{description}

By framing this multi-layered framework around projected trace inclusion and probabilistic measure, we achieve a strict separation of concerns. The first two layers ensure structural correctness: any trace generated by the simulation is structurally constrained to be a valid real-time trace, which in turn projects onto a safe logical trace ($\mathcal{T}_{ST} \upharpoonright_{C_{RT}} = \mathcal{T}_{RT} \upharpoonright_{C_L} \subseteq \mathcal{T}_L$). 

Once the functional safety is guaranteed by this inclusion, the stochastic layer acts as an evaluative lens. By introducing operational probability distributions (e.g., derived from prototype implementation measurements), engineers can perform runtime simulations to quantify performance characteristics--such as the probability distribution of end-to-end reaction times across functional chains--without risking the generation of structurally invalid or unsafe system behaviours.
Said differently, \rtccsl refines by restricting admissible traces; the stochastic layer refines knowledge by introducing a probability measure over the same admissible traces.
This includes effective complementary refinement, in the case of the assignment of the probability measure to zero for specific traces. 

\begin{AEBSproperty}[\textbf{\ccsl Safety Properties}]
\label{subsec:safe-prop}

\begin{table}[t]
\centering
\caption{Logical timing requirements of the AEBS}
\label{tab:aebsccsl}
\begin{tabular}{|l|l|}
\hline
Category & \ccsl constraints \\
\hline
Task execution &
\begin{tabular}[c]{@{}l@{}}
$s_\mathrm{start} \strictalternates s_\mathrm{finish}$ \\
$c_\mathrm{start} \strictalternates c_\mathrm{finish}$ \\
$a_\mathrm{start} \strictalternates a_\mathrm{finish}$
\end{tabular}
\\
\hline
Communication &
\begin{tabular}[c]{@{}l@{}}
$s_\mathrm{finish} \alternates c_\mathrm{receive\_data}$ \\
$c_\mathrm{finish} \alternates a_\mathrm{receive\_data}$
\end{tabular}
\\
\hline
Data consistency &
\begin{tabular}[c]{@{}l@{}}
$c_\mathrm{start} = \sampledon{c_\mathrm{receive\_data}}{c_\mathrm{start}}$ \\
$a_\mathrm{receive\_data} \preccurlyeq (a_\mathrm{start} \;\$\; 1)$ \\
$a_\mathrm{start} \preccurlyeq (a_\mathrm{receive\_data} \;\$\; 1)$ \\
$a_\mathrm{receive\_data} \preccurlyeq a_\mathrm{start}$
\end{tabular}
\\
\hline
\end{tabular}
\end{table}

Following the proposed refinement process, we use \ccsl to express the safety constraints that must be ensured in a modelled system.

Considering the AEBS depicted in \Cref{sec:running}, the logical specification is built around the clocks representing the events of each component: sensor (\(s_\mathrm{start}\), \(s_\mathrm{finish}\)), controller (\(c_\mathrm{receive\_data}\), \(c_\mathrm{start}\), \(c_\mathrm{finish}\)), and actuator (\(a_\mathrm{receive\_data}\), \(a_\mathrm{start}\), \(a_\mathrm{finish}\)).
We show the constructed \ccsl specification in \Cref{tab:aebsccsl}, where we specify three classes of safety requirements.

First, each task is non-reentrant: a new execution cannot start before the previous one has completed. This is expressed as $\strictalternates$ relation between the start and finish clocks.
Second, the communication from sensor to controller (similarly, from controller to actuator) is assumed lossless and ordered. Data produced is received before any subsequent transmission can occur, and expressed as \( \alternates \) relation.
Finally, the specification constrains the relation between data reception and task execution. The controller samples the most recent sensor information at each execution, ensuring that every execution is based on freshly received data. Similarly, actuator executions are kept consistent with the reception of controller outputs, both preventing command loss and guaranteeing their eventual consumption.

\end{AEBSproperty}

\subsection{Real-Time \ccsl}
\label{sec:rtccsl}

The Real-Time Clock Constraint Specification Language (\rtccsl)~\cite{10.1007/978-3-031-63790-2_24,tokariev_2024dec} constitutes the first refinement layer of our timing specification framework. While \ccsl specifies safety requirements as relations between logical instants, \rtccsl introduces additional constraints reflecting implementation aspects of the system and constrains the interpretation of these logical instants in continuous physical time, while allowing timing uncertainty to be expressed through bounded real-time quantities. When the resulting \rtccsl specification is free from deadlock, these bounded quantities define admissible timing budgets under which the original safety properties remain satisfied. At this refinement level, timing analyses can already be performed and timing budgets are typically selected to guarantee global system properties such as end-to-end reaction-time requirements.

For the refinement process developed in this paper, we revisit \rtccsl in two ways. First, timing uncertainty is no longer embedded directly into individual timing constraints but represented explicitly through numerical sequences that can be constrained independently. Second, bounded timing quantities are interpreted as admissible domains rather than merely existential timing constraints. In other words, every value within the specified bounds of the timing quantities must preserve satisfiability of the specification; otherwise the corresponding timing budget is considered invalid. These modifications provide the abstraction layer required for the stochastic refinement introduced in \Cref{sec:stochastic}.

Formally, this language extension supplements the \ccsl time structure \(\langle I, \equiv_I, \prec_I \rangle\) with a labelling function $\rtlabel : I \to \R$ that assigns a real-valued timestamp to ticks.
The labelling function has to be consistent with the time structure, and thus is required to be monotonic over the preorder \(\preccurlyeq_I\):
\begin{equation}
\label{eq:monotone-label}
\begin{aligned}
\forall i,j \in I: \; 
    &i \preccurlyeq_I j \implies \\ \big( &i \in \dom{\rtlabel} \land j \in \dom{\rtlabel} \implies \rtlabel(i) \leq \rtlabel(j) \big)
\end{aligned}
\end{equation}
This property thus implies that the elements of the same equivalence class, defined by \( \equiv_I \), are assigned the same label and vice versa, when their labelling exists.

\rtccsl constraints do not merely assign timestamps to existing ticks. They also impose consistency between the progression of physical time and the existence of logical instants. Informally, if a real-time constraint requires the occurrence of a tick at a given physical time and global time horizon reaches that instant, then the corresponding logical tick must exist. 
This consistency condition is implicitly assumed in the definitions that follow.
Formally, we define the inclusion in the time horizon as $e \leq \top^\rtlabel_I \defeq \exists x \in I: e \leq \rtlabel(x)$, where $\top^\rtlabel_I$ represents the upper bound of the labels \(\rtlabel\) on $I$. 
Thus, whenever an equality of the form $\rtlabel(c[i])=\mathrm{expr}$ appears in a real-time constraint, it is interpreted as $(i \in \dom{\lc{c}} \implies \rtlabel(\lc{c}[i]) = \mathrm{expr}) \land (\mathrm{expr} \leq \top^\rtlabel_I \implies i \in \dom{\lc{c}})$; thereby coupling timestamp assignment and tick existence.

We define the numerical sequences, used as arguments to the real-time constraints, as 
functions from indices to some numerical domain \(X\), meaning \( S_X = \N \to X \) is a set of all sequences over \( X \).
Consequently, the sequences used in the constraints are denoted as \( S_\R \), \( S_{\R_{\geq 0}} \) or \( S_{\R_{> 0}} \) respectively, and reflect part of the correctness conditions associated with the arguments.
We use the same indexing notation as logical clocks, i.e. for \( s \in S_X, i \in \N \), we define \( s[i] \defeq s(i) \).

To express the bounds of the uncertainty, we introduce constant relations on elements of numeric sequences, which constrain the sequences on the per element basis; formally, \(\forall X, a \in S_X, \beta \in X, \Join \in \{=,<,\leq,>,\geq\}\):
\begin{equation}
    a \Join \beta \defeq \forall i \in \N: a[i] \Join_X \beta
\end{equation}
The meaning of \( \Join \) is contextual: on the left of the definition it is a syntactic constraint symbol, on the right it is the corresponding numeric relation \( \Join_X \subseteq X \times X \).

The constant relations over sequence \( s \in S_X \) effectively reduce the definition domain \( X \) to some subset \( Y \subseteq X \).
We thus define the correct usage of the numerical sequences as independent timing budgets by real-time constraints as such that if there exists a trace \( \tau \in \mathcal{T}_{RT} \) that uses the value \(x \in Y \) from the sequence \( x=s[i]\), then there should exist a trace \( \tau' \in \mathcal{T}_{RT} \) with a corresponding sequence \(s' \in S_X\) that uses a different value \( y=s'[i] \in Y, y \neq x \), for the same index in the numerical sequence, for any \( i \in \N \), unless \( Y = \{x\}\).

\rtccsl introduces three new constraints: real-time delay, real-time periodic with jitter and with drift.
Real-time delay can be viewed as a quantitative refinement of causality constraint as it restricts the time difference between the ticks of the argument clocks to concrete durations.
The durations are set to be equal precisely to the elements of nonnegative-valued sequence \( d \in S_{\R_{\geq 0}} \).
The constraint is denoted as \( \lc{b} = \rtdelay{\lc{a}}{d} \), where \(\lc{a}, \lc{b} \in C \),  and is defined as:
\begin{equation}\label{redef:rtdelay}
  \forall i \in \N: \rtlabel(\lc{b}[i]) = \rtlabel(\lc{a}[i]) + d[i]
\end{equation}

Real-time periodic constraint with jitter, \( \lc{r} = \absosc{p}{j}{\varphi} \), and real-time periodic constraint with drift, \( \lc{r} = \relosc{p}{j}{\varphi} \), are constraints that express two patterns of periodicity in real-time.
The periodic constraint with jitter defines a clock whose labels are relative to the true periodicity of period $p$ with offset $\varphi$ up to the error defined by the sequence $j$.
Meanwhile, the periodic constraint with drift defines a clock the next tick of which is set relative to the previous tick, with the period $p$ and the error from the sequence $j$, resulting in accumulation of the errors.
Given the parameters \(  \lc{r} \in C, p \in \R_{> 0}, j \in S_\R, \varphi \in \R_{\geq 0} \), the real-time periodic constraint with jitter is defined as:
\begin{equation}\label{redef:rtperiodic1}
  \forall i \in \N:
  \begin{cases}
    \rtlabel(\lc{r}[i]) = \varphi & \text{if } i = 0\\
    \rtlabel(\lc{r}[i]) = p \cdot i + \varphi + j[i-1] & \text{otherwise}
  \end{cases}
\end{equation}
The real-time constraint with drift is defined respectively as:
\begin{equation}\label{redef:rtperiodic2}
    \forall i \in \N:
  \begin{cases}
    \rtlabel(\lc{r}[i]) = \varphi & \text{if } i = 0\\
    \rtlabel(\lc{r}[i]) = \rtlabel(r[i-1]) + p + j[i-1] & \text{otherwise}
  \end{cases}
\end{equation}
We additionally restrict the sequence of $j$ to values that do not violate the logical clock definition, as the condition: \( \forall i \in \N: j[i+1] > j[i] - p \).

\begin{AEBSproperty}[\textbf{AEBS Real-Time Specification}]
\label{subsec:rt-prop}
Using \rtccsl we are able to express some implementation details of the running example (\Cref{tab:aebsrtccsl}), which are controlled by eight numerical sequences representing execution times (\(s_\mathrm{exec}\), \(c_\mathrm{exec}\), \(a_\mathrm{exec}\)), activation (\(s_\mathrm{jitter}\), \(c_\mathrm{jitter}\), \(a_\mathrm{trig}\)) and communication delays (\(c_{s\to c}\), \(a_{c\to a}\)). 
In this example, the timing bounds are set such that every admissible value within the specified bounds preserves the worst-case end-to-end reaction time of \(\qty{30}{ms}\).\footnote{Reaction time is not part of a specification, as it is an emerging behaviour derived from the system behaviour, and not an independent time budget.}
First, we characterize the uncertain component execution time, establishing admissible execution-time budgets.
Second, we specify component activation rules.
The sensor and controller operate periodically, while the actuator is triggered upon receiving data from the controller. The uncertainty of the periodic components is characterized by jitters. Additionally, a time offset is specified for the controller to avoid initial starvation.
Finally, we constrain the communication delays between the components to be less than 1ms.
\end{AEBSproperty}

\begin{table}[]
    \centering
    \caption{Real-time requirements of the AEBS}
    \label{tab:aebsrtccsl}
    \begin{tabular}{|p{1.2cm}|l|}
        \hline
         Category & \rtccsl constraints \\
         \hline
         Execution & 
            \begin{tabular}[c]{@{}l@{}}
            $\qty{0}{ms} < s_\mathrm{exec} \leq \qty{2}{ms}$ \\
            $s_\mathrm{finish} = \rtdelay{s_\mathrm{start}}{s_\mathrm{exec}}$ \\
            $\qty{0}{ms} < c_\mathrm{exec} \leq \qty{15}{ms}$ \\
            $c_\mathrm{finish} = \rtdelay{c_\mathrm{start}}{c_\mathrm{exec}}$ \\
            $\qty{0}{ms} < a_\mathrm{exec} \leq \qty{3}{ms}$ \\
            $a_\mathrm{finish} = \rtdelay{a_\mathrm{start}}{a_\mathrm{exec}}$ \\
            \end{tabular} \\ \hline
         Activation & 
            \begin{tabular}[c]{@{}l@{}}
            $\qty{-1}{ms} \leq s_\mathrm{jitter} \leq \qty{1}{ms}$ \\
            $s_\mathrm{start} = \absosc{\qty{5}{ms}}{s_\mathrm{jitter}}{\qty{0}{ms}}$ \\
            $\qty{-1}{ms} \leq c_\mathrm{jitter} \leq \qty{1}{ms}$ \\
            $c_\mathrm{start} = \absosc{\qty{10}{ms}}{c_\mathrm{jitter}}{\qty{3}{ms}}$ \\
            $\qty{0}{ms} < a_\mathrm{trig} \leq \qty{1}{ms}$ \\
            $a_\mathrm{start} = \rtdelay{a_\mathrm{receive\_data}}{a_\mathrm{trig}}$ \\
            
            \end{tabular} \\ \hline
         Communi\-cation &  
            \begin{tabular}[c]{@{}l@{}}
            $\qty{0}{ms} < c_\mathrm{s\to c} \leq \qty{1}{ms}$ \\
            $c_\mathrm{receive\_data}  = \rtdelay{s_\mathrm{finish}}{c_\mathrm{s\to c} }$ \\
            $\qty{0}{ms} < a_\mathrm{c\to a} \leq \qty{1}{ms}$ \\
            $a_\mathrm{receive\_data}  = \rtdelay{c_\mathrm{finish}}{a_\mathrm{c\to a} }$ \\
            \end{tabular}\\ \hline
    \end{tabular}
\end{table}
\subsection{Stochastic Extension}
\label{sec:stochastic}

We extend \rtccsl further to express stochastic timing uncertainty in real-time behaviour. Unlike the logical and real-time refinement layers, which constrain the set of admissible traces through projected trace inclusion (\( \mathcal{T}_{RT} \upharpoonright_{C_L} \subseteq \mathcal{T}_L \)), the stochastic extension does not further restrict behavioural possibilities. Instead, it defines a probability measure \(\mu\) over \( \mathcal{T}_{RT}\). Consequently, the set of admissible traces remains unchanged, while the probability measure characterises how likely different executions are to occur during operation or simulation.

We define this probability measure by attaching probability distributions to numerical sequences through stochastic annotations,
which specify how values are sampled from those admissible domains (for instance during simulation). Consequently, stochastic annotations do not introduce new behavioural restrictions; instead, they provide a sampling policy that determines the relative likelihood of timing valuations and, therefore, of the resulting traces.
In practice, the \rtccsl bounds and stochastic distributions may originate from the engineering expertise, manufacturer specifications, or measurements collected from prototype implementations and deployed systems.

As a language construct, the annotations are denoted as \(\texttt{distribute}\; v \;\texttt{as}\; D\), where \(v \in S_X \) is a sequence and \( D \) is some distribution over \( X \), for example, normal or exponential.
Formally, the annotation is defined as \(\forall i \in \N: v[i] \sim D\).
In the current work, successive samples are assumed independent.




\begin{AEBSproperty}[\textbf{AEBS Stochastic Specification}]
\label{subsec:st-prop}

Finally, we refine the description of the AEBS with stochastic annotations, attached to the numerical sequences defined in Specification~\ref{subsec:rt-prop}.
For this purpose, we use a single distribution pattern: a normal distribution truncated to an interval.
We denote it as \( \mathcal{N}(\mu, \sigma)_{[a,b]}\), where \( \mu,\sigma,a,b \in \R\) are mean, standard deviation, lower and upper bounds respectively.
The distribution parameters and bounds remain compatible with the admissible timing bounds defined in the \rtccsl specification. Note that some of them (like the controller execution time) occupy only a subset of the admissible domain.
\end{AEBSproperty}

\begin{table}[]
    \centering
    \caption{Stochastic specification of the AEBS}
    \label{tab:aebsstccsl}
    \begin{tabular}{|p{1.2cm}|l|}
        \hline
         Category & Stochastic annotations \\
         \hline
         Execution & 
            \begin{tabular}[c]{@{}l@{}}
            \( \texttt{distribute}\; s_\mathrm{exec} \;\texttt{as}\; \mathcal{N}(\qty{1.5}{ms}, \qty{0.25}{ms})_{[\qty{0.5}{ms}, \qty{2}{ms}]} \)\\
            \( \texttt{distribute}\; c_\mathrm{exec} \;\texttt{as}\; \mathcal{N}(\qty{3}{ms}, \qty{3}{ms})_{[\qty{1}{ms}, \qty{7}{ms}]} \)\\
            \( \texttt{distribute}\; a_\mathrm{exec} \;\texttt{as}\; \mathcal{N}(\qty{1.2}{ms}, \qty{0.25}{ms})_{[\qty{0.5}{ms}, \qty{2}{ms}]} \)\\
            \end{tabular} \\ \hline
         Activation & 
            \begin{tabular}[c]{@{}l@{}}
            \( \texttt{distribute}\; s_\mathrm{jitter} \;\texttt{as}\; \mathcal{N}(\qty{0}{ms}, \qty{0.5}{ms})_{[\qty{-0.7}{ms}, \qty{0.7}{ms}]} \)\\
            \( \texttt{distribute}\; c_\mathrm{jitter} \;\texttt{as}\; \mathcal{N}(\qty{0}{ms}, \qty{0.5}{ms})_{[\qty{-0.7}{ms}, \qty{0.7}{ms}]} \)\\
            \( \texttt{distribute}\; a_\mathrm{trig} \;\texttt{as}\; \mathcal{N}(\qty{0.6}{ms}, \qty{0.1}{ms})_{[\qty{0.2}{ms}, \qty{0.7}{ms}]} \)\\
            \end{tabular} \\ \hline
         Communi\-cation &  
            \begin{tabular}[c]{@{}l@{}}
            \( \texttt{distribute}\; c_\mathrm{s\to c} \;\texttt{as}\; \mathcal{N}(\qty{0.6}{ms}, \qty{0.1}{ms})_{[\qty{0.2}{ms}, \qty{0.7}{ms}]} \)\\
            \( \texttt{distribute}\; a_\mathrm{c\to a} \;\texttt{as}\; \mathcal{N}(\qty{0.6}{ms}, \qty{0.1}{ms})_{[\qty{0.2}{ms}, \qty{0.7}{ms}]} \)\\
            \end{tabular}\\ \hline
    \end{tabular}
\end{table}

\section{Evaluation}
\label{sec:evaluation}

While the \ccsl constraints defined in Specification~\ref{subsec:safe-prop} capture safety properties, other typical properties concern system performance and are expressed in terms of emergent timing behaviour. One important example is end-to-end reaction time, which is commonly analysed through functional chains~\cite{oueslati.etal_2019,gunzel_2024}. A functional chain identifies a causally related sequence of sensing, computation, communication, and actuation events that links an external stimulus to the corresponding system response.

We evaluate both specifications through a Monte Carlo simulation, generating $10^7$ trace steps in total. For the \rtccsl specification (Specification~\ref{subsec:rt-prop}), admissible sequence values are sampled uniformly within their specified bounds, providing an unbiased exploration of the admissible timing domain. For the stochastic specification (Specification~\ref{subsec:st-prop}), values are sampled according to the annotated probability distributions. Since both specifications induce the same admissible trace space, the comparison highlights the effect of the probability measure rather than differences in behavioural possibilities.

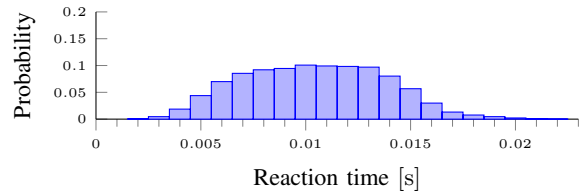
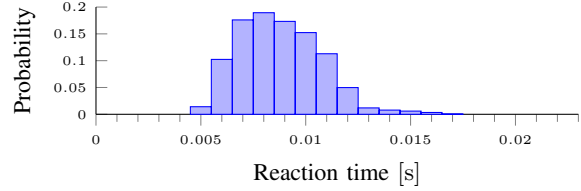
\begin{figure}
    \centering
    \begin{subfigure}[b]{\linewidth}
        \pgfplotsset{%
          width=0.9\linewidth,
          height=3cm,
          scaled y ticks=false,
          yticklabel style={
            font=\tiny,
            /pgf/number format/fixed,
            /pgf/number format/precision=5
          },
          scaled x ticks=false,
          xticklabel style={
            font=\tiny,
            /pgf/number format/fixed,
            /pgf/number format/precision=5
          },
          minor x tick num=4
        }
        \begin{tikzpicture}
          \begin{axis}[use units, ybar, bar width=0.001, axis lines*=left, ymin=0, ymax=0.20,xmin=0,xmax=0.023,xtick distance=0.005,ytick distance=0.05, extra x tick style={grid=major},ylabel={Probability}, xlabel={Reaction time},x unit=s,x label style={font=\small},y label style={font=\small}]
            \addplot table [draw=blue, x=bin, y=total, col sep=comma] {./real.csv};
          \end{axis}
        \end{tikzpicture}
        \caption{Uniformly-explored AEBS specification}
        \label{fig:real-reaction}
    \end{subfigure}\\%
    \begin{subfigure}[b]{\linewidth}
        \pgfplotsset{%
          width=0.9\linewidth,
          height=3cm,
          scaled y ticks=false,
          yticklabel style={
            font=\tiny,
            /pgf/number format/fixed,
            /pgf/number format/precision=5
          },
          scaled x ticks=false,
          xticklabel style={
            font=\tiny,
            /pgf/number format/fixed,
            /pgf/number format/precision=5
          },
          minor x tick num=4
        }
        \begin{tikzpicture}
          \begin{axis}[use units, ybar, bar width=0.001, axis lines*=left, ymin=0, ymax=0.20,xmin=0,xmax=0.023,xtick distance=0.005,ytick distance=0.05, extra y tick style={grid=major}, extra x tick style={grid=major},ylabel={Probability}, xlabel={Reaction time},x unit=s,x label style={font=\small},y label style={font=\small}]
            \addplot table [draw=blue, x=bin, y=total, col sep=comma] {./stochastic.csv};
          \end{axis}
        \end{tikzpicture}
        \caption{Stochastically-annotated AEBS specification}
        \label{fig:stochastic-reaction}
    \end{subfigure}
    \caption{Reaction time distributions of the running example}
    \label{fig:reactions}
\end{figure}

From the obtained traces we derive the functional chains following the first-to-first semantics~\cite{Feiertag2008ACF}.
The resulting end-to-end reaction times are aggregated into the empirical distributions shown in \Cref{fig:real-reaction,fig:stochastic-reaction}.
Both specifications satisfy the \(<\qty{30}{ms}\) reaction-time requirement. This is expected, since the admissible timing domains defined by the \rtccsl specification were selected to guarantee the correctness for all admissible valuations.
While the uniformly explored \rtccsl specification represents all admissible executions as equally likely, the stochastic specification reveals the timing behaviour expected during operation.
Such information is particularly valuable when evaluating software updates, comparing alternative implementations, or estimating the timing margin available for future functionality additions without modifying the underlying correctness guarantees.

The complete experimental setup, implemented in OCaml, is available at \url{https://zenodo.org/records/20580040}.

\section{Discussion} 
\label{sec:discussion}

Traditional logical-time and synchronous modelling approaches provide an effective abstraction for reasoning about system behaviour by hiding implementation-level timing effects behind logical instants and synchronisation assumptions \cite{benveniste2003,lee2016introduction}. While this abstraction is valuable during early design, it is less amenable to refinement because many logical synchronisations must eventually be realised through execution, communication, and scheduling delays. Rather than deriving conditions under which synchrony remains valid, our approach preserves logical requirements while progressively introducing explicit timing quantities that can later be constrained and characterised statistically.

A second challenge concerns the derivation of component-level timing budgets from system-level requirements. As in classical timing analysis and functional-chain engineering \cite{Feiertag2008ACF,oueslati.etal_2019,gunzel_2024}, multiple allocations of execution, communication, and activation delays may satisfy the same end-to-end requirement. The \rtccsl layer therefore does not eliminate timing-budget engineering but makes the resulting assumptions explicit and traceable throughout the refinement process.

Finally, the stochastic extension should be viewed as a refinement of operational knowledge rather than of correctness. Correctness is established at the \rtccsl level through admissible timing domains. Stochastic annotations merely characterise how executions are distributed within these domains. This separation allows timing distributions to evolve as new measurements become available, for example after software updates, deployment on different hardware platforms, or collection of operational data, without requiring redevelopment of the underlying specification.
This supports continuous timing assessment throughout the system lifecycle rather than restricting timing analysis to a single design phase.

\section{Conclusion}
\label{sec:conclusion}

We have successfully developed and implemented a refinement-oriented timing specification framework based on \ccsl. The framework supports the progressive refinement of timing knowledge from logical requirements, to admissible real-time budgets, and finally to stochastic operational models that quantify the likelihood of emergent timing properties such as reaction times.

The future work will focus on formalising the probabilistic semantics of the stochastic layer, enabling statistical guarantees on simulation results and application of advanced Monte Carlo techniques for faster reaction-time estimation~\cite{degenhardt.etal_2025feb}.

\bibliographystyle{IEEEtran}
\bibliography{ref}

\end{document}